\documentclass[aps,prd,reprint,superscriptaddress, nofootinbib]{revtex4-1}
\usepackage{amsmath}
\usepackage{amsfonts}
\usepackage{amssymb}
\usepackage{graphicx,color}
\usepackage{bm}
\usepackage{braket}										
\usepackage{subcaption}
\usepackage[caption=false]{subfig}

\renewcommand{\Re}{\text{Re}}

\renewcommand{\d}{d}

\newcommand{\be}{\begin{equation}}
\newcommand{\ee}{\end{equation}}
\newcommand{\Tr}[1]{\text{Tr}\left[{#1}\right]}
\newcommand{\exv}[1]{\langle {#1} \rangle}
\newcommand{\pc}[1]{{\psi_{{#1}}^\text{cl}}}
\newcommand{\pcd}[1]{{\psi_{{#1}}^{\text{cl}\dagger}}}
\newcommand{\pq}[1]{{\psi_{{#1}}^\text{q}}}
\newcommand{\pqd}[1]{{\psi_{{#1}}^{\text{q}\dagger}}}

\graphicspath{
	{gfx/vq_un/}
	{gfx/vq_tr/}
	{gfx/crit_para/}
	{gfx/coll_un/}
	{gfx/coll_tr/}
}

\begin{document}

\title{Self-similar Evaporation and Collapse in the Quantum Portrait of Black Holes}

\author{Valentino F.\ Foit}
\email[]{foit@nyu.edu}
\affiliation{Department of Physics, New York University, 4 Washington Place, New York, NY 10003, USA}
\affiliation{Physik Department, Technische Universit\"{a}t M\"{u}nchen, James Franck Str.\ 1, 85747 Garching, Germany}
\author{Nico Wintergerst}
\email[]{nico.wintergerst@fysik.su.se}
\affiliation{Arnold Sommerfeld Center, Ludwig-Maximilians-Universit\"{a}t, Theresienstr.\ 37, 80333 M\"{u}nchen, Germany}
\affiliation{The Oskar Klein Centre for Cosmoparticle Physics,
Department of Physics, Stockholm University,
AlbaNova, 106 91 Stockholm, Sweden}
\date{\today}

\begin{abstract}
We investigate Hawking evaporation in a recently suggested picture in which black holes are Bose condensates of gravitons at a quantum critical point.
There, evaporation of a black hole is due to two intertwined effects. Coherent excitation of a tachyonic breathing mode is responsible for the collapse of the condensate, while incoherent scattering of gravitons leads to Hawking radiation. To explore this, we consider a toy model of a single bosonic degree of freedom with derivative self-interactions. We consider the real-time evolution of a condensate and derive evaporation laws for two possible decay mechanisms in the Schwinger-Keldysh formalism. We show that semiclassical results can be reproduced if the decay is due to an effective two-body process, while the existence of a three-body channel would imply very short lifetimes for the condensate. In either case, we uncover the existence of scaling solutions in which the condensate is at a critical point throughout the collapse. In the case of a two-body decay we moreover discover solutions
that exhibit the kind of instability that was recently conjectured to be responsible for fast scrambling.

\end{abstract}

\maketitle

\section{Introduction}

Black hole (BH) solutions in general relativity (GR) do not possess global timelike Killing vector fields. As a consequence, black holes of mass $M$ emit radiation \cite{Hawking:1974rv, *Hawking:1974sw}, semiclassically found to obey a thermal\footnote{Of course, due to the decrease of the black hole mass, the observed spectrum will not be thermal for a finite mass black hole. Exact thermality is a consequence of the semiclassical limit corresponding to black hole mass $M \to \infty$, Planck mass $M_p \to \infty$, and Schwarzschild radius $R_s = M/M_p^2$ fixed.} distribution with a temperature $T_H \propto 1/M$. 
In a semiclassical analysis, an evaporating black hole can turn pure states into mixed states \cite{Hawking:1976ra} - the evaporation process appears to violate the unitarity principle of quantum mechanics.

More recent advances, most prominently in the holographic \cite{adscft_largen, *adscft_witten} understanding of black holes have led to the widespread belief that black hole evaporation is after all a unitary process. In the above terms, this necessarily implies a breakdown of the semiclassical description at some point during the evaporation process, most likely around the halfway point of evaporation when the black hole is still large \cite{Page:1993df, *Page:1993wv}. In particular, this calls for an extension of the semiclassical approximation towards a more complete picture.

In this paper we develop a microscopic mechanism responsible for particle loss of black holes within the condensate picture that has been proposed in \cite{Dvali:2011aa, *Dvali:2012gb, *Dvali:2012rt, Dvali:2012en}.
There, black holes are described as multiparticle quantum states of gravitons at a point of \emph{collective} strong coupling, or quantum criticality. The key assumption is that all relevant physics originates in collective effects of gravitons whose wavelength is given by the characteristic length scale $\ell$ of the gravitational background. This in principle allows a quantum description of objects of size $\ell \gg \ell_P$, with $\ell_P = \sqrt{\hbar G}$ the Planck length, within the low energy approximation to the effective action of gravity, i.e.\ the Einstein-Hilbert action. 
Within this description, two effects can lead to significant deviations from semiclassical predictions. 
First, the criticality of the condensate may ultimately lead to the appearance of almost gapless modes, corresponding to collective excitations of the graviton condensate. Their presence can lead to strong \emph{quantumness} even for large black holes \cite{Flassig:2012re}.
Second, corrections suppressed by the effective number of gravitons may become important on sufficiently long timescales.   
This, in turn, serves to restore unitarity in the evaporation process \cite{Dvali:2012en}.

In the case of black holes, wavelengths of the order of the Schwarzschild radius $R_s$ are expected to dominate. Under this assumption, black holes indeed lie at a point of collective strong coupling. This may be the microscopic origin of black hole entropy \cite{Dvali:2012en} and the underlying reason for black hole quantumness \cite{Flassig:2012re} and scrambling \cite{Dvali:2013vxa}.

The strength of graviton-graviton interactions is given by perturbative general relativity as
\begin{equation} \label{alpha}
	\alpha = \frac{\ell_P^2}{\ell^2},
\end{equation}
where $\ell$ is the characteristic wavelength of the interacting gravitons.
For large systems, the particles are extremely weakly coupled. 
However, due to their bosonic nature, they can occupy states with a large number of particles. In a Hartree picture, each graviton sees a collective binding potential produced by the other gravitons. For sufficiently large numbers, this allows for the formation of \emph{self-sustained} bound states.
The number of constituents $N$ of such a bound state can be estimated using the virial theorem \cite{Dvali:2011aa}, where self-sustainability is achieved 
for $\alpha N = 1$. The average wavelength of a graviton in the system is related to $N$ via  Eq.\ \eqref{alpha}
\begin{equation} \label{bhsize}
	\ell = \ell_P\sqrt{N},
\end{equation}
while the total mass of the bound state is approximately given by the sum of the energies of the individual constituents $M = N \frac{\hbar}{\ell} = \sqrt{N} \frac{\hbar}{\ell_P}$. As a consequence, $N$ becomes the universal characteristic of the condensate.

Black hole formation from gravitational collapse may be understood as bringing the condensate to the critical point \cite{Dvali:2012en}. In fact, $N$ is set by the energy stored in the gravitational field, roughly constant throughout the collapse. On the other hand $\alpha$ increases until black hole formation sets in for $\alpha N \simeq 1$. 
The formation of BH in ultrahigh energy scattering at center of mass energies $\sqrt{s} \gg M_p$ also appears accessible within this approach. 
Graviton numbers of order $1/\alpha \sim s/M_p^2$ are seen to dominate scattering amplitudes, either as intermediate states in $2 \to 2$ processes \cite{Kuhnel:2014xga}, or as final states in $2 \to N$ scattering \cite{Dvali:2014ila}. 

Condensate decay, on the other hand, is due to the interactions between the constituents. In essence, incoherent scattering between condensed gravitons leads to ejection from the condensate and a change of particle number $N \to N'$. Concurrent to the emission, the black hole collapses and readjusts its size: $R_s \to R_s' = \ell_p \sqrt{N'}$. Consequently, the graviton-graviton coupling is given by $\alpha = \ell_p^2 / R_s'^2 = 1/N'$. After one emission cycle, the condensate is still at a point of collective strong coupling with lower particle number and stronger coupling.

In this paper we address the mechanism for the evaporation of the condensate in a microscopic picture (see e.g.\ \cite{Casadio:2014vja, Kuhnel:2014zja} for other considerations towards Hawking evaporation in this framework). We consider a toy model with an interaction structure that bears important similarities with GR.
We demonstrate that the evaporation of the condensate is to leading order due to scattering of two condensed particles. We further show that if one of the two participants can rescatter into the condensate, the obtained rate greatly exceeds the rate that is expected from semiclassical considerations. If, on the other hand, this is disallowed, as for example for a homogeneous condensate, we obtain an evaporation rate that scales as
\begin{equation}
	\dot{N} = - \frac{1}{\ell_P \sqrt{N}} + \mathcal{O}(N^{-3/2}),
	\label{eq:hawking}
\end{equation}
or, using $M \sim M_p\sqrt{N}$, in terms of the bound state mass
\begin{equation}
	\dot{M} = - \frac{M_p^4}{M^2} + \mathcal{O}(N^{-2}).
\end{equation}
In the large $N$ limit, this corresponds to the mass loss that is obtained semiclassically for a black hole of temperature $T_H = \frac{\hbar}{\ell_P \sqrt{N}}$ \cite{Dvali:2011aa}. 
If this picture is applicable to gravity, unitarity is obviously never an issue; the entire black hole can radiate away without any loss of information. The difference to semiclassical results is encoded in
$1/N$ corrections.

Moreover, we discover that in the case of a two-body decay, the critical solutions exhibit an instability with Lyapunov coefficient $\lambda \sim 1/\ell_P \sqrt{N}$. This implies that along these solutions, entanglement is generated on a timescale $R_s \log R_s$. It has been conjectured \cite{Dvali:2012en} that this is the microscopic origin for the fast scrambling property of black holes \cite{Hayden:2007cs,*Sekino:2008he}.

Our paper is organized as follows.
We discuss our assumptions in section \ref{proto}. In sections \ref{keldysh} and \ref{effaction} we describe the basic physical mechanism responsible for the evaporation of the condensate. We derive the equations governing the collapse of the condensate in section \ref{collapse} and take evaporation into account in section \ref{subsec:evap}. The final section then contains discussions on the solutions found in case of three- (\ref{sec:threebody}) and two-body (\ref{sec:twobody}) decay.
From here on, the speed of light $c$ and Planck's constant $\hbar$ are chosen to be unity.

\section{From gravity to prototype\label{proto}}
In the condensate picture for black holes, the dynamics of collapse and Hawking evaporation are due to two intertwined effects.
The coherent excitation of a tachyonic breathing mode of the condensate leads to collapse of the black hole. This is a process involving only the gravitons of the condensate.
At the same time, incoherent scattering allows for the production of  gravitons that can escape the black hole.
In principle, the former may be accounted for through mean field evolution, while the latter is due to the interaction of the mean field with quantum fluctuations.

In a gauge where the linear\footnote{As usual, we linearize around Minkowski. This also gives us a preferred time slicing.} graviton obeys the relations
\begin{equation}
h_{0i} = 0,\, h^\mu_\mu = 0,\, \partial^\mu h_{\mu\nu} = 0,
\label{eq:lin_gauge}
\end{equation}
the corresponding time evolution is generated by a Hamiltonian that in Fourier space takes the (suggestive) form
\begin{multline}
\hat{H} = \int d^3k \sum_{\lambda = 1,2} |{\bf k}| \hat{a}_{{\bf k},\lambda}^\dagger \hat{a}_{{\bf k},\lambda} 
\pm \sum_n M_p^{2-n}\times\\
\int d^3k_1 \ldots d^3k_n 
\frac{{\bf k}_i {\bf k}_j} {\sqrt{\prod_{l=1}^n |{\bf k}_l|}}
 P^{(n)}\left(\hat{a}\right)\, \delta^{(3)}\left(\sum {\bf k}_i\right).
\label{eq:grav_ham}
\end{multline}
Here $\lambda = 1,2$ corresponds to the two transverse polarizations of the graviton and the $a_{{\bf k},\lambda}^\dagger, a_{{\bf k},\lambda}$ are creation and annihilation operators of gravitons with polarization $\lambda$. The functions $P^{(n)}$ comprise all possible degree $n$ monomials of said creation and annihilation operators, thereby generating the infinite series of vertices present in an interacting massless spin-2 theory. The interaction term will generically contain also the longitudinal and temporal polarizations of the graviton, depending on the choice of nonlinear gauge. 

In our picture, a black hole roughly corresponds to a set of quantum states $|\text{BH}\rangle$ in the interacting theory with a large occupation number of gravitons in a single mode $\hat{a}_\text{BH} = \int d^3k \alpha_k \hat{a}_k$. Note that the $\hat{a}_k$ comprise annihilation operators of all possible polarizations, in order for the black hole state to be part of the physical spectrum.

In principle, (\ref{eq:grav_ham}) contains all the relevant information for the analysis of the non-equilibrium behavior of the states $|\text{BH}\rangle$. However, the corresponding vertices, stemming from a Poincar\'e invariant field theory, will not conserve particle number. Moreover, we are dealing with the presence of the infinite series of vertices and the ambiguity due to gauge redundancy. An explicit treatment of this theory is extremely difficult. Therefore, we are in dire need of simplifying assumptions.

For now, these will be:
\begin{itemize}
\item[(i)] Reduce the number of polarizations to a single mode, thereby also removing the gauge ambiguity.
\item[(ii)] Focus on particle number conserving processes.
\item[(iii)] Assume that the relevant dynamics of the condensate is captured already by the lowest order interaction. Due to (ii) we consider only the quartic vertex.
\end{itemize}
We will briefly comment on their viability in the Appendix.

The latter two assumptions will prevent us from learning anything about the actual spectrum of emitted particles. Nevertheless, we will see that they already allow for very interesting conclusions on the condensate dynamics. Take note here that due to (ii) and (iii), all momenta involved in the leading order collision processes are expected to be of the same order. We will therefore also neglect the momentum dependent prefactor of the quartic interaction term.

With these assumptions, we arrive at the Hamiltonian
\begin{multline}
\hat{H} = \int d^3k |{\bf k}| \hat{a}_{{\bf k}}^\dagger \hat{a}_{{\bf k}} \\
- M_p^{-2}
\int d^3k_1 \ldots d^3k_4 \hat{a}_{{\bf k}_1}^\dagger \hat{a}_{{\bf k}_2}^\dagger \hat{a}_{{\bf k}_3} \hat{a}_{{\bf k}_4}\delta^{(3)}\left(\sum {\bf k}_i\right).
\label{eq:ham_fin}
\end{multline}
Under our assumptions, the difference to previous prototype models for graviton condensates \cite{Dvali:2012en,Flassig:2012re,Dvali:2013vxa} reduces to the relativistic dispersion relation. We will see, however, that it is precisely this feature that is responsible for interesting properties. Note also that our Hamiltonian is based on derivative interactions. The difference in the interaction as compared to \cite{berkhahn_classicalons} is due to the inherently nonrelativistic nature of the model considered there.

Before we start analyzing the dynamics of (\ref{eq:ham_fin}), we utter a word of caution. Of course, the simplified Hamiltonian is void of quite a few important features of gravitation. Besides the simplifications in terms of number conservation, we have eliminated the longitudinal and temporal modes from the dynamics. In GR, it is precisely these modes that are responsible for the gravitational potential. Most simply, their presence may be modeled through the inclusion of a trapping potential. This will be left for future work. Let us also note that the reduction to particle number conserving processes simplifies the structure of the interacting vacuum. This will allow us for example to decompose correlators in a coherent state basis without having to worry about subtleties. Implications of the nontrivial nature of the interacting vacuum on the bound state description of black holes have been studied in \cite{Hofmann:2014jya, *Gruending:2014rja}.

\section{Schwinger-Keldysh formalism\label{keldysh}}
The following section provides an elementary review of the Schwinger-Keldysh \cite{schwingerkeldysh1,*schwingerkeldysh2} formalism that will allow for a proper treatment of the real time dynamics of collapse and evaporation.
It may be skipped by the experienced reader. Comprehensive introductions may be found e.g.\ in \cite{kamenev2011field}. Our analysis presents a generalization of previous results (found e.g. in \cite{stoof1, stoof_duine_prl, stoof_duine_pra} and many subsequent works) to systems with a relativistic dispersion relation.

We are interested in the time evolution of an unstable condensate which is initially\footnote{A precise study of the dynamics of black hole formation in the condensate picture is an interesting issue on its own. Here we simply assume the presence of an initial condensate.} described by a (normalized) density matrix $\hat\rho(t_i)$. The expectation value of any observable $\hat{\cal{O}}$ is given by
\begin{align}
\exv{\cal{O}} (t) &= \Tr{U(t_i,t) \hat{\mathcal{O}} U(t,t_i)\hat\rho} \\ &= \Tr{U(t_i,t_f) U(t_f,t) \hat{\mathcal{O}} U(t,t_i)\hat\rho} ,
\end{align}
where $U(t_1, t_2) = \exp{(-i H (t_1 - t_2))}$ is the time evolution operator. The second equation has been obtained through an insertion of $U(t_i,t_f)U(t_f,t_i)$ for $t_f$ in the asymptotic future. It has served to extend the integration path from $t_i \to t \to t_i$ to $t_i \to t_f \to t_i$, the so-called Keldysh contour, which we will denote by ${\cal C}$.

With use of the Keldysh contour, expectation values can be obtained from a generating functional through the introduction of corresponding sources into the \mbox{Hamiltonian}
\begin{align}
H^\pm_J &= H \pm J(t) \hat{\mathcal{O}}\,, \\
Z[J] &= \Tr{U_J({\cal C})\hat\rho}\,, \label{eq:gen_fun}\\
\exv{\mathcal{O}}(t) &= \frac{i}{2}\left.\frac{\delta Z[J]}{\delta J(t)}\right|_{J=0}, \,t_i \le t \le t_f \,.
\end{align}
Here $H^+$ and $H^-$ are the Hamilton operators along the forward and backward contour, respectively.

As usual, (\ref{eq:gen_fun}) may be turned into a path integral by introducing at each timestep an appropriate partition of unity. In this case, we use coherent states that are eigenstates of the annihilation operators appearing in (\ref{eq:ham_fin}), $\hat{a}_{\bf k}\ket{\psi} = \psi_{\bf k} \ket{\psi}$. One obtains
\begin{equation}
Z = \int {\cal D}\psi {\cal D}\psi^\dagger e^{i S[\psi,\psi^\dagger]}\,,
\end{equation}
where the ``action'' $S$ is given by
\begin{multline}
S[\psi,\psi^\dagger] = \int_{\cal C} dt \int d^3k_1 \Bigg\{ i \psi_{\bf k_1}^\dagger \partial_t \psi_{\bf k_1} - |{\bf k_1}| \psi_{\bf k_1}^\dagger \psi_{\bf k_1} \\
+ M_p^{-2}\int d^3k_2 \ldots d^3k_4 \psi_{\bf k_1}^\dagger \psi_{\bf k_2}^\dagger \psi_{\bf k_3} \psi_{\bf k_4}\delta^{(3)}\left(\sum {\bf k_i}\right) \Bigg\}. \label{eq:action}
\end{multline}
The information on the initial state is encoded in the correlation of the field on the forward and backward branch.

The time integral in (\ref{eq:action}) may be brought into conventional form by introducing forward and backward fields $\psi^\pm$ that live on the forward (backward) branch of the Keldysh contour. Performing a so called Keldysh rotation by introducing the ``classical'' and ``quantum'' fields 
\begin{equation}
\pc{} = \frac{1}{\sqrt{2}}\left(\psi^+ + \psi^-\right)\,,\, \pq{} = \frac{1}{\sqrt{2}}\left(\psi^+ - \psi^-\right)\,,
\end{equation}
one obtains
\begin{multline}
S[\pc{} ,\pcd{},\pq{},\pqd{}] = \\
 \int dt\, d^3{\bf k}_1 d^3{\bf k}_2\Bigg\{ i \vec{\psi}_{\bf k_1}^\dagger {\cal K}({\bf k_1}, {\bf k_2}) \vec{\psi}_{\bf k_2} \\
+ M_p^{-2}\int d^3{\bf k}_3 d^3{\bf k}_4  \delta^{(3)}\left(\sum {\bf k_i}\right) \\
\times\left(\pcd{\bf k_1} \pcd{\bf k_2} \pc{\bf k_3} \pq{\bf k_4} + \pqd{\bf k_1} \pqd{\bf k_2} \pq{\bf k_3} \pc{\bf k_4}\right) \Bigg\}
 + \text{h.c.} \,.
\label{eq:action_sk}
\end{multline}
Here we have introduced $\vec{\psi}_{\bf k} \equiv (\pc{\bf k}, \pq{\bf k})^T$ and the kinetic matrix ${\cal K}$ is defined as 
\begin{multline}
\label{eq:kmat}
{\cal K}({\bf k}_1, {\bf k}_2) \equiv \\
\begin{pmatrix}
0 & \delta({\bf k}_1 - {\bf k}_2)(i \partial_t - |{\bf k}_1|) \\
\delta({\bf k}_1 - {\bf k}_2)(i \partial_t - |{\bf k}_1|) & \Sigma_K\left({\bf k}_1, {\bf k}_2\right)
\end{pmatrix}\,.
\end{multline}
We have introduced the Keldysh self-energy $\Sigma_K$, whose precise value depends on the interactions and is of no particular interest to us. Its presence, however, is important, since it contains the information on the correlators of the forward and backward fields and thereby on the initial density matrix.
Note that at this point, one may equivalently seek a formulation for the action (\ref{eq:action_sk}) in terms of real fields
 by translating $\Psi$ and $\Psi^\dagger$ into a real scalar field and its canonical momentum. However, our focus on particle number conserving processes is more straightforwardly implemented in the current language.

The classical mean-field dynamics of the condensate arise from (\ref{eq:action_sk}) as the solution to the saddle-point equations that has $\pq{} = 0$. In this case, variation of (\ref{eq:action_sk}) with respect to $\pqd{}$ yields the Gross-Pitaevskii equation for the classical field. This is the equation of motion that describes the mean-field dynamics of the condensate:
\begin{multline}
\label{eq:GP}
i \partial_t \pc{\bf k} = |{\bf k}|\pc{\bf k} \\- M_p^{-2} \int d^3{\bf k}_1 d^3{\bf k}_2 d^3{\bf k}_3 \delta\left({\bf k} + {\bf k}_1 - {\bf k}_2 - {\bf k}_3 \right) \\
\times \pcd{\bf k_1} \pc{\bf k_2} \pc{\bf k_3}\,.
\end{multline}
For a condensate with attractive self-interactions, the normalized solutions to this equation will correspond to a collapsing condensate of $N$ particles once $N$ surpasses a critical value \cite{ruprecht_collapse}\footnote{Even for smaller $N$, the condensate is only metastable and can always decay through macroscopic tunneling \cite{stoof_collapse}.}.

\section{Effective action\label{effaction}}
The evaporation of the condensate is due to scattering of condensed particles into the quasi-particle cloud. These effects may be taken into account most readily by integrating out the quasi-particles. To this regard, we separate the fields into a  condensate part and fluctuations.
\begin{equation}
\pc{} = \phi + \delta\varphi\, , \,\, \pq{} = \phi^q + \delta\varphi^q\,.
\label{eq:bg_fluc}
\end{equation}
Inserting (\ref{eq:bg_fluc}) into (\ref{eq:action_sk}) gives rise to a plethora of terms, of which only few have a relevant effect. The reason for this lies in the fact that we are dealing with a condensate with $N \gg 1$ at $T = 0$. The quasiparticle occupation is much lower than that of the condensate mode.
We may therefore focus on terms that are at most quadratic in the fluctuations. All corrections that arise from higher order terms are proportional to the density of fluctuations and are thus at least $1/N$-suppressed.

Primarily, we are interested in the loss of condensed particles due to incoherent scattering. Two processes contribute:

\begin{itemize}
\item[(i)] Two-body decay: This corresponds to scattering of two condensed particles and subsequent emission of two quasiparticles. In the action the relevant operator is quadratic in the fluctuations.
\item[(ii)] Three-body decay: Again, two condensed particles scatter. Here, however, only one of the outgoing particles leaves the condensate. The corresponding operator is linear in the fluctuations.
\end{itemize}

Note that the three-body process is only non-vanishing for an inhomogeneous condensate; otherwise it is incompatible with the conservation of momentum.

Following our above remarks, we construct the following expression for the quasiparticle action:
\begin{widetext}
\begin{multline}
S_\text{qp} = \int dt\, d^3{\bf k}_1 d^3{\bf k}_2\Bigg\{ i \vec{\delta\phi}_{\bf k_1}^\dagger {\cal K}_\text{qp}({\bf k_1}, {\bf k_2}) \vec{\delta\phi}_{\bf k_2} + M_p^{-2}\int d^3{\bf k}_3 d^3{\bf k}_4  \delta^{(3)}\left(\sum {\bf k_i}\right) \\
\times
\left(2 \delta\phi^{cl\dagger}_{\bf k_1} \phi_{{\bf k_2}}^\dagger \phi_{{\bf k_3}} \phi^{q}_{\bf k_4} + \delta\phi^{cl}_{\bf k_1} \phi_{{\bf k_2}}^\dagger \phi_{{\bf k_3}}^\dagger \phi^{q}_{\bf k_4} + \delta\phi^{q}_{\bf k_1} \phi_{{\bf k_2}}^\dagger \phi_{{\bf k_3}}^\dagger \phi_{{\bf k_4}}\right) \Bigg\}
 + \text{h.c.} \,,
\end{multline}
\end{widetext}

We have here introduced the quasiparticle vector $\vec{\delta\phi} \equiv \left(\delta\phi^{cl},\delta\phi^{q},\delta\phi^{cl\dagger},\delta\phi^{q\dagger}\right)^T$, as well as  the quadratic quasiparticle operator ${\cal K}_\text{qp}$, which reads
\begin{equation}
{\cal K}_\text{qp} = 
\begin{pmatrix}
{\cal K} + {\cal A} & {\cal B} \\
{\cal B}^\dagger & {\cal K}^\dagger + {\cal A}
\end{pmatrix}\,,
\end{equation}
with ${\cal K}$ defined in Eq.\ \eqref{eq:kmat} and (suppressing momentum labels and integrals)
\begin{align}
{\cal A} &= 
2 M_p^{-2}\begin{pmatrix}
\phi^{cl\dagger}\phi^q & \phi^{cl\dagger}\phi^{cl} \\
\phi^{cl\dagger}\phi^{cl} & \phi^{cl\dagger}\phi^q
\end{pmatrix} + \text{h.c.}\,,\\
{\cal B} &= 
M_p^{-2}\begin{pmatrix}
\phi^{cl}\phi^q & \frac{1}{2}\phi^{cl}\phi^{cl} \\
\frac{1}{2}\phi^{cl}\phi^{cl} & \phi^{cl}\phi^q
\end{pmatrix}\,.
\end{align}
We have retained only terms that are at maximum linear in the quantum field $\phi^q$. A loss term in the Gross-Pitaevskii equation can only originate from a term in the effective action that is linear in $\phi^q$; all other effects that may alter the condensate dynamics are $1/N$ suppressed. Note that the quadratic part of the fluctuation action allows one to read off the quasiparticle spectrum; diagonalization leads to the celebrated Bogoliubov modes \cite{bogolyubov}. 

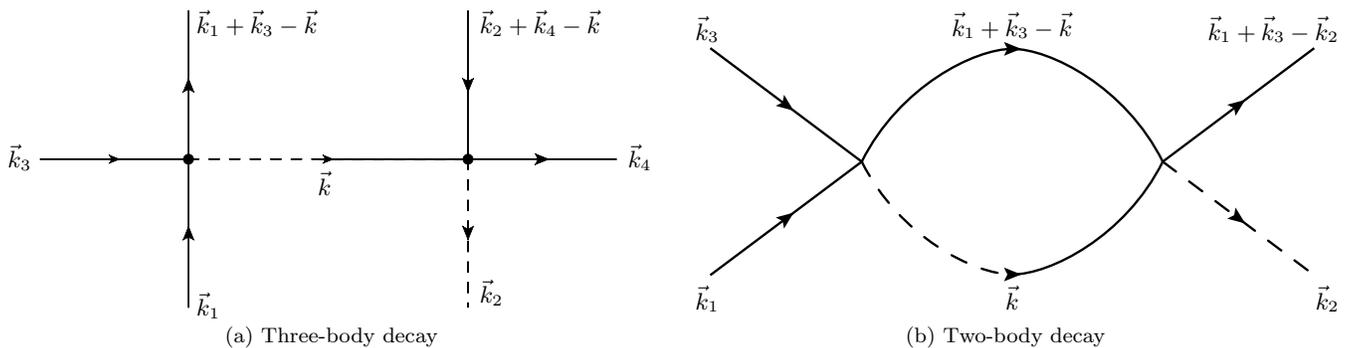
\begin{figure*}[t]
\setlength{\unitlength}{\linewidth}
\begin{minipage}[b]{.5\linewidth}
\begin{picture}(1,0.22)%
    \put(0.15,0.0){\makebox(0,0)[l]{\strut{}$\vec{k}_1$}}
    \put(0.01,0.11){\makebox(0,0)[l]{\strut{}$\vec{k}_3$}}
    \put(0.15,0.21){\makebox(0,0)[l]{\strut{}$\vec{k}_1 + \vec{k}_3 - \vec{k}$}}
    \put(0.24,0.09){\makebox(0,0)[l]{\strut{}$\vec{k}$}}
    \put(0.47,0.11){\makebox(0,0)[l]{\strut{}$\vec{k}_4$}}
    \put(0.36,0.21){\makebox(0,0)[l]{\strut{}$\vec{k}_2 + \vec{k}_4 - \vec{k}$}}
    \put(0.36,0.01){\makebox(0,0)[l]{\strut{}$\vec{k}_2$}}
    \put(0,0.0){\includegraphics[width=\columnwidth]{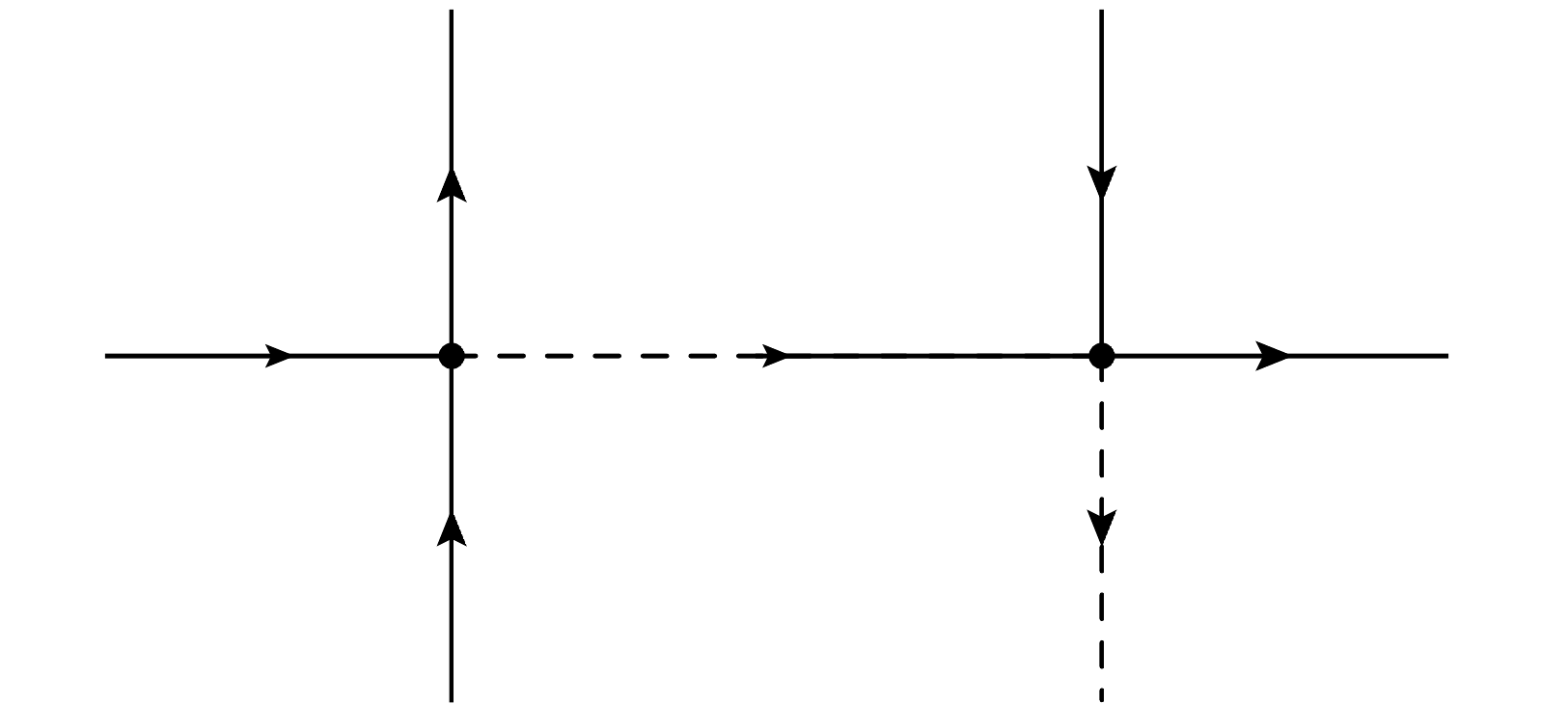}}
\end{picture}
\subcaption{Three-body decay}\label{fig:threebody}
\end{minipage}%
\begin{minipage}[b]{.5\linewidth}
\begin{picture}(1,0.22)
    \put(0.02,0.005){\makebox(0,0)[l]{\strut{}$\vec{k}_1$}}
    \put(0.02,0.205){\makebox(0,0)[l]{\strut{}$\vec{k}_3$}}
    \put(0.25,0.005){\makebox(0,0)[l]{\strut{}$\vec{k}$}}
    \put(0.21,0.208){\makebox(0,0)[l]{\strut{}$\vec{k}_1 + \vec{k}_3 - \vec{k}$}}    
    \put(0.48,0.005){\makebox(0,0)[l]{\strut{}$\vec{k}_2$}}
    \put(0.40,0.205){\makebox(0,0)[l]{\strut{}$\vec{k}_1 + \vec{k}_3 - \vec{k}_2$}}
    \put(0.03,0.02){\includegraphics[width=0.9\columnwidth]{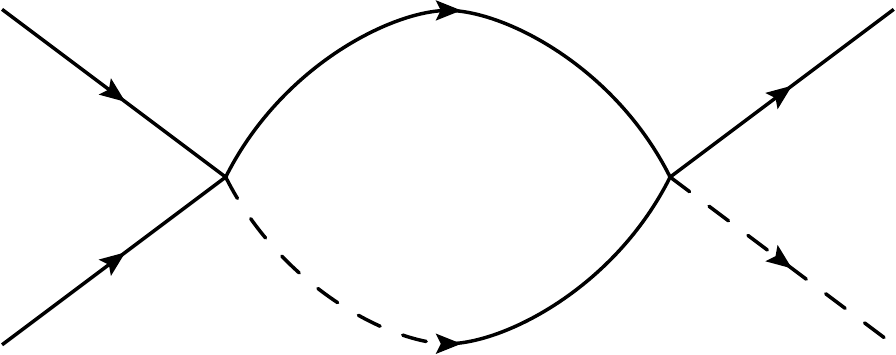}}%
  \end{picture}%
\subcaption{Two-body decay}\label{fig:twobody}
\end{minipage}

\caption{Lowest order diagrams responsible for the imaginary part of the self-energy. The external lines correspond to condensed particles while internal lines correspond to quasiparticles; solid lines identify classical and dashed lines quantum fields. The imaginary parts are induced when in (a) $\vec{k}$ and in (b) $\vec{k}$ and $\vec{k}_1 + \vec{k}_2 - \vec{k}$  go on-shell.}
	\label{fig:manybodyscat}
\end{figure*}

In order to obtain a modified Gross-Pitaevskii equation that incorporates the effects of the quasiparticles, we integrate out the latter. Loss terms will be generated by the 
diagrams shown in Fig.\ \ref{fig:manybodyscat}, which obtain imaginary parts due to on-shell fluctuations. Fig.\ \ref{fig:threebody} describes the three-body decay process \cite{stoof_duine_prl}, while Fig.\ \ref{fig:twobody} corresponds to the two-body decay. 
In the effective action, we generate the contribution
\begin{equation}
\delta S = \frac{i}{M_p^4} \int dt d^3{\bf k}_1 d^3{\bf k}_2 \Gamma({\bf k}_1, {\bf k}_2) \left(\phi_{{\bf k}_1}\phi^{q\dagger}_{{\bf k}_2} - \phi^{q}_{{\bf k}_1}\phi^{\dagger}_{{\bf k}_2}\right)\,, 
\end{equation}
where we have introduced
\begin{equation}
\Gamma({\bf k}_1, {\bf k}_2) \equiv \Gamma^{(3)}({\bf k}_1, {\bf k}_2) + \Gamma^{(2)}({\bf k}_1, {\bf k}_2)
\end{equation}
and defined
\begin{align}
\Gamma^{(3)}({\bf k}_1, {\bf k}_2) &\equiv \nonumber \\
\int d^3{\bf k} d^3{\bf k}_3& d^3{\bf k}_4 \delta\left(|{\bf k}_1|+|{\bf k}_2|-|{\bf k}|-|{\bf k}_1 + {\bf k}_2 - {\bf k}|\right) 
\label{eq:threebodydecay}
\nonumber \\
&\times \phi^\dagger_{{\bf k}_1+{\bf k}_3-{\bf k}}\phi_{{\bf k}_3}\phi^\dagger_{{\bf k}_4}\phi_{{\bf k}_2+{\bf k}_4-{\bf k}} \,,\\
\Gamma^{(2)}({\bf k}_1, {\bf k}_2) &\equiv \nonumber \\
\int d^3{\bf k} d^3{\bf k}_3& \delta\left(|{\bf k}_1|+|{\bf k}_2|-|{\bf k}|-|{\bf k}_1 + {\bf k}_2 - {\bf k}|\right) \nonumber \\
&\times \phi^\dagger_{{\bf k}_1 + {\bf k}_2 - {\bf k}_3} \phi_{{\bf k}_3}\,.
\label{eq:twobodydecay}
\end{align}
Here the former contribution is due to diagram \ref{fig:threebody}, giving rise to a loss term in the effective action that is proportional to six powers in the fields, hence scaling with the particle number as $N^3$. On the other hand, the contribution from Fig.\ \ref{fig:twobody} is proportional to only four powers in the fields, scaling as $N^2$. Note that expression (\ref{eq:twobodydecay}) is obtained from Fig.\ \ref{fig:twobody} by evaluating the integral $\int dk^0 \delta(k^0 - |{\bf k}|)$, following from the on shell contribution from the propagator of ${\bf k}$. Let us here also mention that one-loop diagrams with more external legs 
correspond to processes in which one or more of the emitted quasiparticles scatter on the condensate. These give rise to contributions which are proportional to the quasiparticle density; they are thus $1/N$-suppressed and will be neglected. Such additional suppression is also present for all higher loop processes.

In order to ensure that we do not erroneously count processes where all particles rescatter into the condensate, we should impose a constraint on the fluctuation momentum ${\bf k}$ and only integrate over momenta outside the support of the condensate. This is vital in particular  when considering a homogeneous condensate, where this ensures that the three-body decay rate vanishes. On the other hand, for an inhomogeneous condensate, this will only lead to a numerical renormalization of the decay rate, unless the window of allowed momenta depends explicitly on $N$. In absence of a compelling argument why this should be the case, there will always be a critical $N$ after which the three-body decay dominates.

Once the above loss terms are included, the Gross-Pitaevskii equation reads
\begin{multline}
\label{eq:last_gp}
i \partial_t \phi_{\bf k} = |{\bf k}|\phi_{\bf k} \\+ M_p^{-2} \int d^3{\bf k}_1 d^3{\bf k}_2 d^3{\bf k}_3 \left({\bf k} + {\bf k}_1 - {\bf k}_2 - {\bf k}_3 \right) \phi^\dagger_{\bf k_1} \phi_{\bf k_2} \phi_{\bf k_3} \\
-i \int d^3{\bf k}_2 \Gamma({\bf k}, {\bf k}_2) \phi_{{\bf k}_2}\,.
\end{multline}
From this, we can immediately read off the change in the number of condensed particles
\begin{equation}
\label{eq:lossrate}
\frac{dN}{dt} = -\int d^3{\bf k}_1 d^3{\bf k}_2 \Gamma({\bf k}_1, {\bf k}_2) \phi^\dagger_{{\bf k}_1}\phi_{{\bf k}_2}
\end{equation}

Having integrated out the quasiparticles,
the final step towards an effective action that describes the collapse and evaporation of our condensate is now to integrate out the quantum field $\phi^q$. 
We can do this via integrating \emph{in} an auxiliary ``noise'' field $\eta_{\bf k}(t)$.
\begin{align}
Z = \int {\cal D}\phi\,{\cal D}\phi^q\,{\cal D}\phi^\dagger\,{\cal D}\phi^{q\dagger}\,{\cal D}\eta\, e^{i S_\eta[\phi,\phi^\dagger]}\,,&\\
S_\eta[\phi,\phi^\dagger] = \int dt d^3{\bf k} \Bigg\{\phi^{q\dagger}_{\bf k} \left({\cal E}_{\bf k}(\phi,\phi^{\dagger}) - \eta_{\bf k}\right) + &\text{h.c.} \nonumber \\
+ \frac{1}{\Sigma_{\bf k}}\eta_k^* \eta_k + {\cal O}\left(\frac{1}{N}\right)&\Bigg\}\,,
\end{align}
where ${\cal E}$ is the operator corresponding to the Gross-Pitaevskii equation (\ref{eq:last_gp}).
Integrating out $\phi^q$ now constrains the classical field $\phi^c$ to obey a Langevin equation with Gaussian noise $\eta$
\begin{equation}
\label{eq:full_gp}
{\cal E}_{\bf k}(\phi,\phi^\dagger) = \eta_{\bf k}\,.
\end{equation}

The dynamics described by (\ref{eq:full_gp}) may in principle be obtained numerically. Instead, we will here take a simpler route and seek a variational solution to the \emph{averaged} Gross-Pitaevskii equation. 
Taking note that the dissipative equation (\ref{eq:full_gp}) does not directly follow from a variational principle, we proceed by dividing the variational approach into two steps:
First, we shall look for a variational solution to the simpler problem without the dissipative term. This will provide us with an equation for the condensate size. Second, we supplement this with Eq.\ (\ref{eq:lossrate}) in order to take into account the loss of particles. There we will focus on two regimes. First, we consider the case in which the three-body decay \eqref{eq:threebodydecay} is allowed. In the second case, the dominant contribution comes from the two-body decay \eqref{eq:twobodydecay}.

\section{Variational approach}

\subsection{Collapse\label{collapse}}

In the following we will focus on the average dynamics of the condensate. By the Gaussian nature of the noise $\eta$, its first moment vanishes, and will hence not influence the dynamics.
Without the dissipative term,
Eq.\ \eqref{eq:full_gp} can be obtained as the stationary point of the following Lagrangian. 
\begin{multline} \label{lag_comp}
\mathcal{L}_{\bf k}
 =  \frac{i}{2} \left(\phi_{\bf k}^* \dot{\phi}_{\bf k} - \phi_{\bf k} \dot{\phi}^*_{\bf k}\right) - |{\bf k}||\phi_{\bf k}|^2
\\ + M_p^{-2}\int d^3{\bf k}_1 d^3{\bf k}_2 d^3{\bf k}_3  \delta^{(3)}\left(\sum {\bf k_i}\right) 
\left(\phi^\dagger_{\bf k} \phi^\dagger_{{\bf k}_1} \phi_{{\bf k}_2} \phi_{{\bf k}_3}\right)\\
 + \text{h.c.}\,,
\end{multline}

We extremize Eq.\ (\ref{lag_comp}) with respect to a set of spherically symmetric trial functions. Guided by simplicity, we choose a Gaussian ansatz in real space\footnote{Note that for a nonrelativistic harmonically trapped condensate, the ground state wavefunction can indeed be well approximated by a Gaussian even in the presence of interactions \cite{PhysRevA.56.1424, stringari_review_1999}.}
\begin{equation} \label{eq:ansatz}
	\phi(r,t) = A(t) \left(\frac{3}{4 R(t)}\right)^{\frac{3}{2}} e^{-\frac{\pi}{2}\left(\frac{3}{4}\right)^2\left(\frac{r^2}{R(t)^2} - i r^2 b(t)\right)}.
\end{equation}
$A(t)$ is the complex amplitude, $R(t)$ the real width of the condensate and $r$ the radial coordinate.
The function $b(t)$ can later be identified with the velocity of the collapse. The normalization of $\phi$ is chosen such that $|A(t)|^2 = N$ and the numerical factors simplify the calculation.

Fourier transforming Eq.~\eqref{eq:ansatz} and inserting the result into Eq.~\eqref{lag_comp},
we obtain the averaged Lagrangian density $L = \int \d^3k \tilde{\mathcal{L}}$
\begin{multline}
\label{eq:varlag}
L = -\frac{i}{2} (A^* \dot{A} - A \dot{A}^*)-\frac{3}{4} |A|^2 R^2 \dot{b} - \\
\frac{3|A|^2}{2 R} \sqrt{1 + b^2 R^4} + \frac{27|A|^4}{128 \sqrt{2} M_p^2 R^3}\,.
\end{multline}
We can now understand the collapse dynamics as a variational problem of the time dependent parameters $s= \{ A, A^*, R, b \}$, which obey the equations of motion 
\be 
\frac{\d}{\d t} \left(\frac{\partial L}{\partial \dot{s}_i}\right) - \frac{\partial L}{\partial s_i} =0.
\ee

The equations of motion for the amplitude simplifies to particle number conservation
\begin{equation} \label{I}	
	\frac{\d}{\d t } |A|^2 = 0,
\end{equation}
or in other words $\dot{N} = 0$. 

Next, we can relate $b$ to the collapse velocity $\dot{R}$ by varying for $b$ and using Eq.~\eqref{I}
\begin{equation} \label{II}
b = \frac{\dot{R}}{R^2\sqrt{1 - \dot{R}^2}}.
\end{equation}
This equation completely determines the function $b$ in terms of $R$ and $\dot{R}$.

The expression for the condensate width is obtained by variation with respect to $R$. After substitution of Eq.~\eqref{I} and \eqref{II}, one obtains
\begin{equation} \label{eq:qddot}
	\ddot{R} = \frac{1}{R^3}\left(1 - \dot{R}^2\right) \left(R^2- \frac{1}{\sqrt{2}}\left(\frac{3}{4}\right)^3\frac{N}{M_p^2} \sqrt{1-\dot{R}^2}\right)\,.
\end{equation}

Note the Lagrangian after integrating out $b$
\begin{equation}
\label{eq:lag_pointpart}
L = -\frac{3 N}{2 R}\sqrt{1-\dot{R}^2} +\frac{1}{\sqrt{2}}\left(\frac{3}{4}\right)^3 \frac{N^2}{M_p^2 R^3}\,.
\end{equation}
We recognize it as the Lagrangian of relativistic point particle with a mass that depends on $R$.

\subsubsection{Slow collapse}
It is instructive to first consider the case of small collapse velocities, as this corresponds to a nonrelativistic limit and allows us to qualitatively compare our expressions with existing results from the literature.

In the limit of small velocities and small acceleration, Eq.\ (\ref{eq:qddot}) reads
\begin{equation} 
\label{eq:qddot_slow}
\ddot{R} \approx \frac{1}{R} + \frac{\dot{R}^2}{2R} - \frac{1}{\sqrt{2}}\left(\frac{3}{4}\right)^3 \frac{N}{M_p^2 R^3}\,.
\end{equation}
The first term in Eq.\ (\ref{eq:qddot_slow}) corresponds to the outward force due to the kinetic energy of the bosons; its scaling with inverse $R$ is dictated by Heisenberg's uncertainty principle. The third term is due to the attractive interactions and can, for sufficiently large $N$, overcome the repulsive force. In that case, the condensate collapses. In comparison with known results in the literature (e.g.\ \cite{PhysRevA.56.1424,stoof_duine_prl}), the second term corresponds to corrections due to the relativistic dispersion relation.

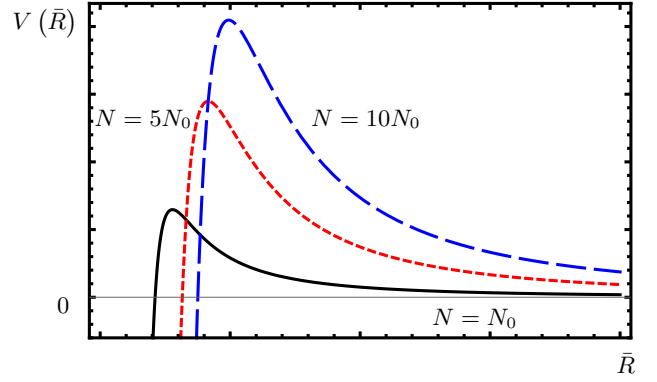
\begin{figure}[t]
  \setlength{\unitlength}{0.93\linewidth}
  \centering
  \begin{picture}(0.95,0.64)%
      \put(0.03,0.11){\makebox(0,0)[r]{\strut{} 0}}%
      \put(-0.01,0.58){\makebox(0,0){\strut{}$V\left(\bar{R}\right)$}}%
      \put(0.95,0.01){\makebox(0,0){\strut{}$\bar{R}$}}%
      \put(0.63,0.09){\makebox(0,0)[l]{\strut{}$N=N_0$}}%
      \put(0.072,0.42){\makebox(0,0)[l]{\strut{}$N=5N_0$}}%
      \put(0.43,0.42){\makebox(0,0)[l]{\strut{}$N=10N_0$}}%
    \put(0.05,0.04){\includegraphics[width=0.85\linewidth]{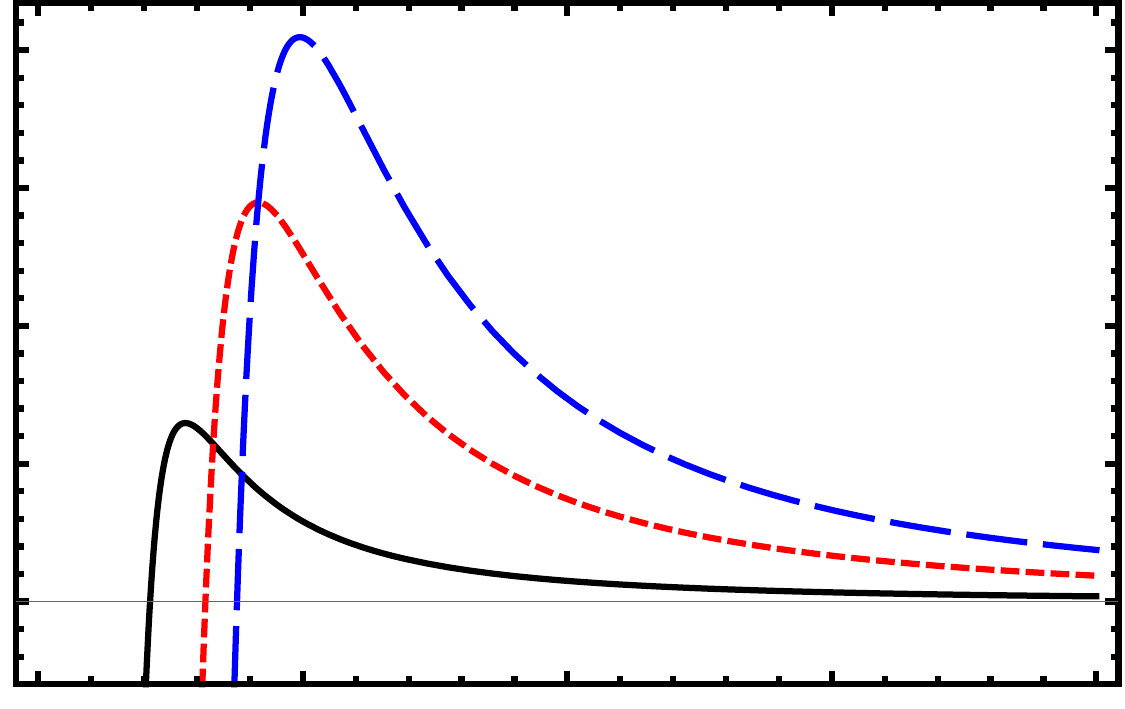}}%
  \end{picture}%
\caption{Effective potential for various values of $N$.}
\label{fig:vq_un}
\end{figure}

In the small velocity limit, the Lagrangian (\ref{eq:lag_pointpart}) may be canonically normalized. The corresponding equation of motion will then give us a simple picture of the time evolution of the width as the motion of a particle $m \ddot{R} = - \frac{\d}{\d R} V(R,N)$ in a one dimensional potential. To see this, let us take the small velocity limit also in the Lagrangian (\ref{eq:lag_pointpart}):
\begin{equation}
L \approx \frac{3}{4} N \frac{\dot{R}^2}{R} - \frac{3 N}{2 R} + \frac{27}{128\sqrt{2}}\frac{N^2}{M_p^2 R^3} \,.
\end{equation}
We may canonically normalize the kinetic term through the redefinition $R = \bar{R}^2$. From the canonical Lagrangian
\begin{equation}
L = 3 N \dot{\bar{R}}^2 - \frac{3 N}{2 \bar{R}^2} + \frac{1}{\sqrt{2}}\left(\frac{3}{4}\right)^3\frac{N^2}{M_p^2 \bar{R}^6}\,,
\end{equation} 
we can conclude that the motion corresponds to that of a particle with mass $m = 6 N$ in the effective potential
\begin{equation}
V(\bar{R}) = \frac{3 N}{2 \bar{R}^2} - \frac{1}{\sqrt{2}}\left(\frac{3}{4}\right)^3\frac{N^2}{M_p^2 \bar{R}^6}\,.
\end{equation}
We plot the effective potential 
in Fig.\ \ref{fig:vq_un}.
The potential $V(\bar{R})$ possesses a maximum, located at 
\begin{equation}
	\bar{R}_+^2 = R_+ \sim \ell_P \sqrt{N}\,.
\end{equation}
It turns out that for $R \sim R_+$, the criticality condition is fulfilled:
\begin{equation}
\alpha N \sim \frac{\ell_P^2}{R^2} N \sim 1 \,.
\end{equation}

Finally note that the runaway behavior for large $R$ is due to the fact that we have not included an external trapping potential for the condensate.

\subsection{Evaporation\label{subsec:evap}}

The decay rate of the condensate according to Eq.\ \eqref{eq:lossrate} 
can be evaluated for a spherically symmetric collapse using the ansatz Eq.\ \eqref{eq:ansatz} which yields 
\begin{equation} \label{eq:ndotm}
	\dot{N} = - \frac{1}{M_p^{4}} \frac{N^2}{R^5} \left( c_{(3)} N + c_{(2)} \right).
\end{equation}
Here the $c_{(i)}$ are dimensionless constants, whose value depends on the precise form of the condensate wavefunction. 
The factors of $M_p$ can be read off straightforwardly from the diagrams Fig.\ \ref{fig:manybodyscat}, while the dependence on $R$ follows from dimensional grounds.

As we have discussed, in our setup both $c_{(3)}$ and $c_{(2)}$ are nonzero. In that case, for sufficiently large $N$ the evaporation will be dominated by the three-body decay with $\dot{N} \sim -N^3$.
However, we admit the possibility that the precise form of the interaction or the form of the wavefunction can effectively close the three-body decay channel.
 In this case, the evaporation will occur with $\dot{N} \sim -N^2$. Both cases exhibit very interesting properties, which we shall explore in the next sections.

Equations (\ref{eq:qddot}) and (\ref{eq:ndotm}) dictate the evolution of the condensate described by the Hamiltonian (\ref{eq:ham_fin}) in the Gaussian approximation, as long as higher order correlators of the fluctuations can be neglected. 

\section{Solutions}
\label{sec:sols}

\subsection{Three body decay}
\label{sec:threebody}

One may solve the collapse and evaporation equations (\ref{eq:qddot}) and (\ref{eq:ndotm}) numerically for generic initial conditions, drawing a complete picture of the behavior of the condensate in the variational approach. However, it turns out that in case of the three-body decay, the equations possess a simple set of analytic collapse solutions: 
\begin{subequations}
\label{eq:coll_sols}
\begin{align}
	R(t) &= R_i - v t \\
	N(t) &= \sqrt{\frac{2 v}{c_{(3)}}}\,\frac{\left(R_i - v t\right)^2}{\ell_P^2}\,.
\end{align}
\end{subequations}
The parameter $v$ is fixed via the algebraic relation
\begin{equation}
\left(1 - v ^2\right) \left(\sqrt{c_{(3)}} - \left(\frac{3}{4}\right)^3 \sqrt{v(1-v^2)}\right)=0.
\end{equation}
Two properties of these solutions are immediately evident. For one, they exhibit self-similarity: The criticality condition $R \sim \ell_P \sqrt{N}$ is fulfilled throughout the collapse. Second, the total evaporation time is proportional to $t_\text{coll} \sim R_i \sim \ell_P \sqrt{N_0}$. Irrespective of the collapse velocity $v$, this immediately rules out a three-body decay of the form (\ref{eq:threebodydecay}) as the mechanism of black hole evaporation. In gravity semiclassical arguments indicate a black hole lifetime $t_\text{BH} \sim M_\text{BH}^3/M_p^4 \sim \ell_P N_0^{3/2}$, which is a factor $N_0$ longer than the lifetime of the condensate considered here. 

\begin{figure*}[t]
\setlength{\unitlength}{\columnwidth}
\begin{minipage}[b]{.5\linewidth}
  \begin{picture}(0.95,0.75)%
      \put(0.04,0.695){\makebox(0,0){\strut{}$v$}}%
      \put(0.9,0.063){\makebox(0,0){\strut{}$c$}}%
      \put(0.0,0.085){\includegraphics[height=0.55\linewidth]{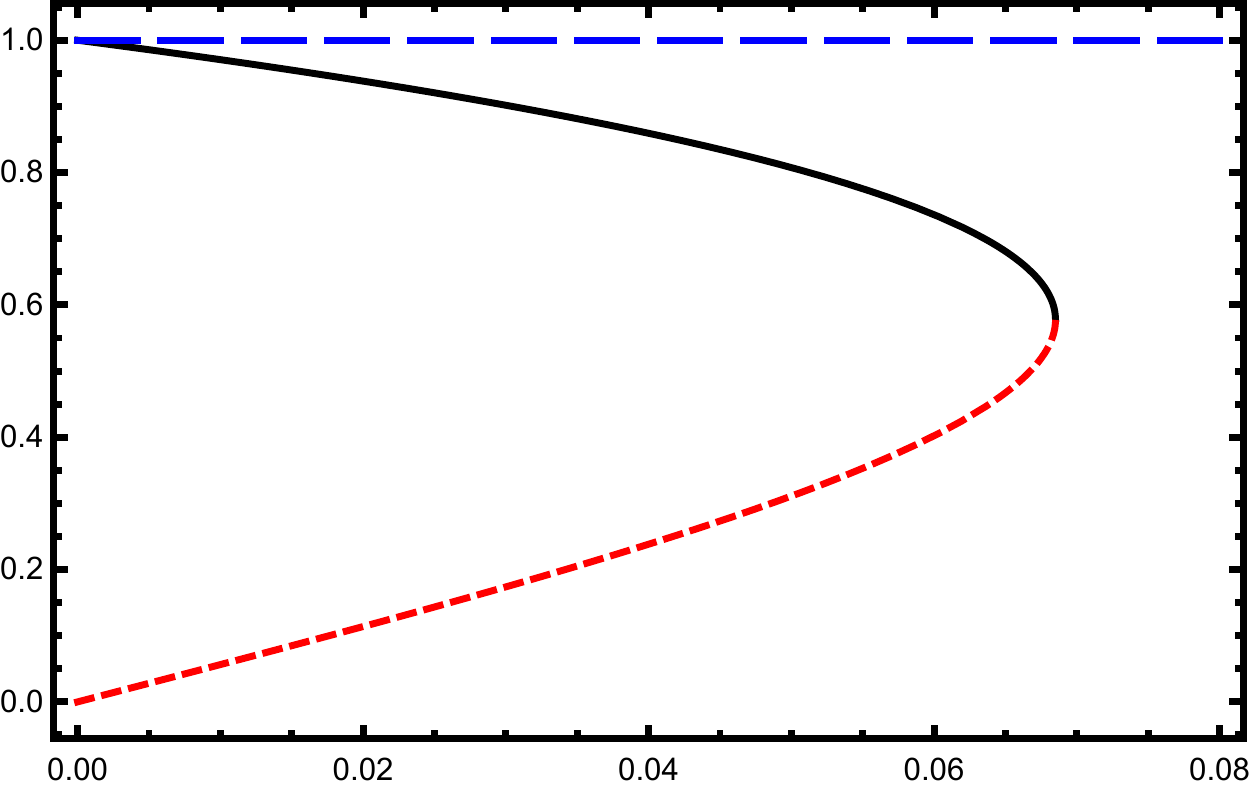}}%
  \end{picture}%
\subcaption{Velocity parameter $v$ as a function of $c$. While the solution $v = 1$ (blue, long-dashed) is $c$-independent, the solutions with $v = v_{2,3}$ (black, solid and red, dashed) are only real for $c \le c_{\text{cr}}$. For $c = c_\text{cr}$, the solutions disappear in a saddle-node bifurcation.}\label{fig:three_body_sols}
\end{minipage}%
\begin{minipage}[b]{.5\linewidth}
  \begin{picture}(0.95,0.65)%
      \put(0.04,0.65){\makebox(0,0){\strut{}$\Re{\lambda_+}\ell_P\sqrt{N}$}}%
      \put(0.9,0.018){\makebox(0,0){\strut{}$c$}}%
      \put(0.0,0.04){\includegraphics[height=0.55\linewidth]{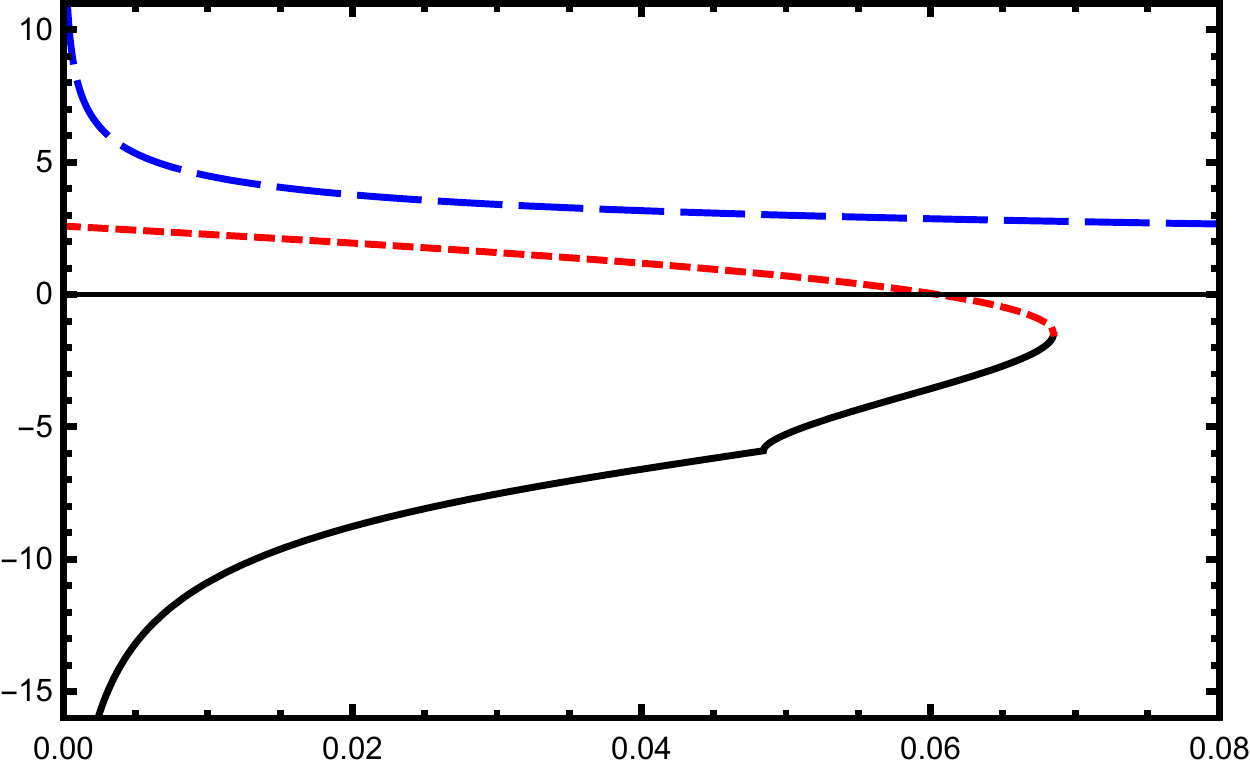}}%
  \end{picture}%
\subcaption{Largest eigenvalue as a function of $c$. The solution $v = 1$ (blue, long-dashed) always possesses an unstable direction; the solution $v = v_{2}$ (black, solid) is absolutely stable, while $v = v_{3}$ (red, short-dashed) is unstable for small $c$. For $c \lesssim  0.05$, $\lambda^{(3)}_\pm$ are degenerate. This is the origin of the kink in the black curve.}\label{fig:three_body_stab}
\end{minipage}

\caption{Solution properties for the three-body decay.}
	\label{fig:three_body_props}
\end{figure*}
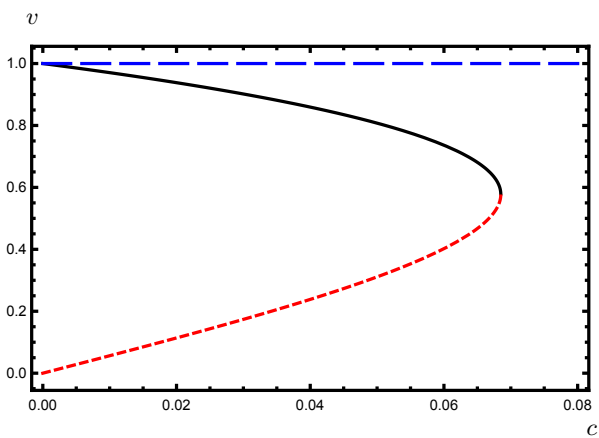
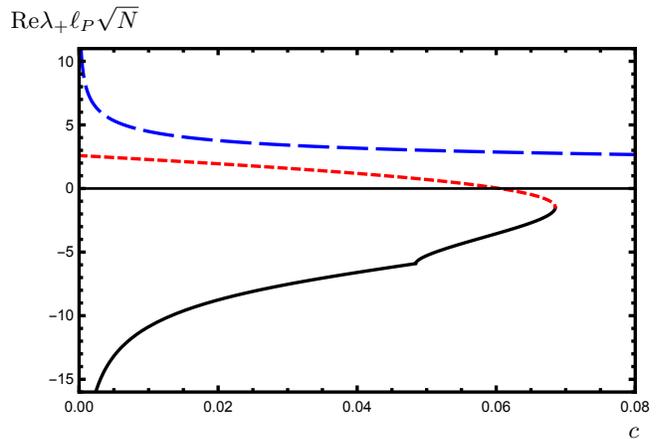

Let us nevertheless comment on some of the interesting features of the solutions (\ref{eq:coll_sols}).
We find three solutions in the allowed range $0 \leq v \leq 1$; a $c$-independent solution $v = 1$ as well as the two $c$-dependent ones\footnote{The corresponding values are given by $v = \frac{1}{\sqrt{3}}\cos\left(f(c)\right) \pm \sin\left(f(c)\right)$, where $f(c) = \frac{1}{3}\mathrm{arctan}\left(\sqrt{\left(\frac{3}{4}\right)^{3}\frac{1}{2c^2}-1}\right)$.}. 
Let us point to the curiosity that the two latter solutions are only real for values of $c_{(3)}$ that are smaller than a critical value $c_\text{crit}$. At $c_{(3)} = c_\text{crit}$ both solutions disappear in a saddle-node bifurcation. We have illustrated this behavior in Fig.\ \ref{fig:three_body_sols}.

Let us now investigate the stability properties of the critical solutions (\ref{eq:coll_sols}).
As we will see, not all three solutions are stable. However, as long as the two $c$-dependent solutions exist, at least one of them presents an attractor.

Since we are only interested in deviations from critical behavior, and not in the stability of a specific solution, it is sufficient to decompose $R$ as
\begin{equation}
R(t) = \left(\frac{c_{(3)}}{2v}\right)^{1/4} \sqrt{N(t)}\ell_P + \delta{R}(t)\,,\\
\end{equation}
and to then combine Eqs.\ (\ref{eq:qddot}) and (\ref{eq:ndotm}) and linearize in $\delta{R}$.
For $v = 1$, we obtain
\begin{align}
\delta\ddot{R} = \frac{5}{\ell_P^2 N}\left(\frac{2}{c_{(3)}}\right)^{\frac{1}{2}}\delta\dot{R} - \frac{3}{\ell_P\sqrt{N}}\left(\frac{2}{c_{(3)}}\right)^{\frac{1}{4}}\delta{R}\,,
\end{align}
while in the other cases, we have
\begin{align}
\delta\ddot{R} = \frac{12 v^2 - 2}{\ell_P^2 N}\left(\frac{2v}{c_{(3)}}\right)^{\frac{1}{2}}\delta\dot{R} - \frac{6v}{\ell_P\sqrt{N}}\left(\frac{2 v}{c_{(3)}}\right)^{\frac{1}{4}}\delta{R}\,.
\end{align}
The eigenvalues of the stability matrix read
\begin{equation}
\lambda_\pm^{(1)} = \frac{-3 \pm \sqrt{29}}{2^{3/4}{c_{(3)}}^{1/4}\ell_P \sqrt{N}}\,
\end{equation}
in the first case and
\begin{equation}
\lambda_\pm^{(2,3)} = \left(\frac{2 v_{2,3}}{c_{(3)}}\right)^{\frac{1}{4}}\cdot\frac{-3 v_{2,3} \pm \sqrt{2 - 3v_{2,3}^2}}{\ell_P \sqrt{N}}\,
\end{equation}
in the latter. The expressions on the corresponding solutions for $v$ are rather lengthy. Important here is only that real parts of the eigenvalues are always negative for $v_2$, while for $v_3$ the larger one is positive for $c$ below some threshold value.
Henceforth, at least one of the two solutions is absolutely stable. The solution $v = 1$, on the other hand, possesses an unstable direction; under small perturbations, it flows towards the solution $v \approx 1-2.8c_{(3)}$. The scaling behavior therefore remains unaltered. We display the stability behavior in Fig.\ \ref{fig:three_body_stab} by plotting the real part of the larger eigenvalue for all three solutions.

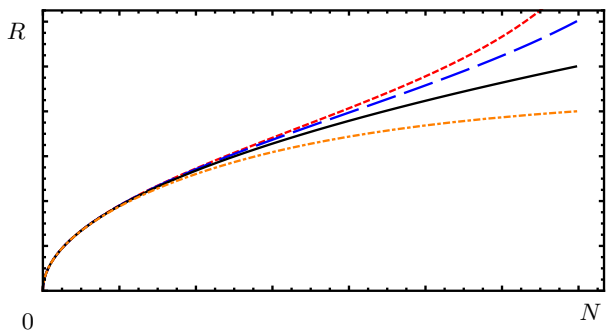
\begin{figure}[t]
  \setlength{\unitlength}{0.93\linewidth}
  \centering
  \begin{picture}(0.95,0.55)%
      \put(0.0,0.0){\makebox(0,0)[r]{\strut{} 0}}%
      \put(-0.03,0.48){\makebox(0,0){\strut{}$R$}}%
      \put(0.92,0.015){\makebox(0,0){\strut{}$N$}}%
      \put(0.0,0.04){\includegraphics[width=0.88\linewidth]{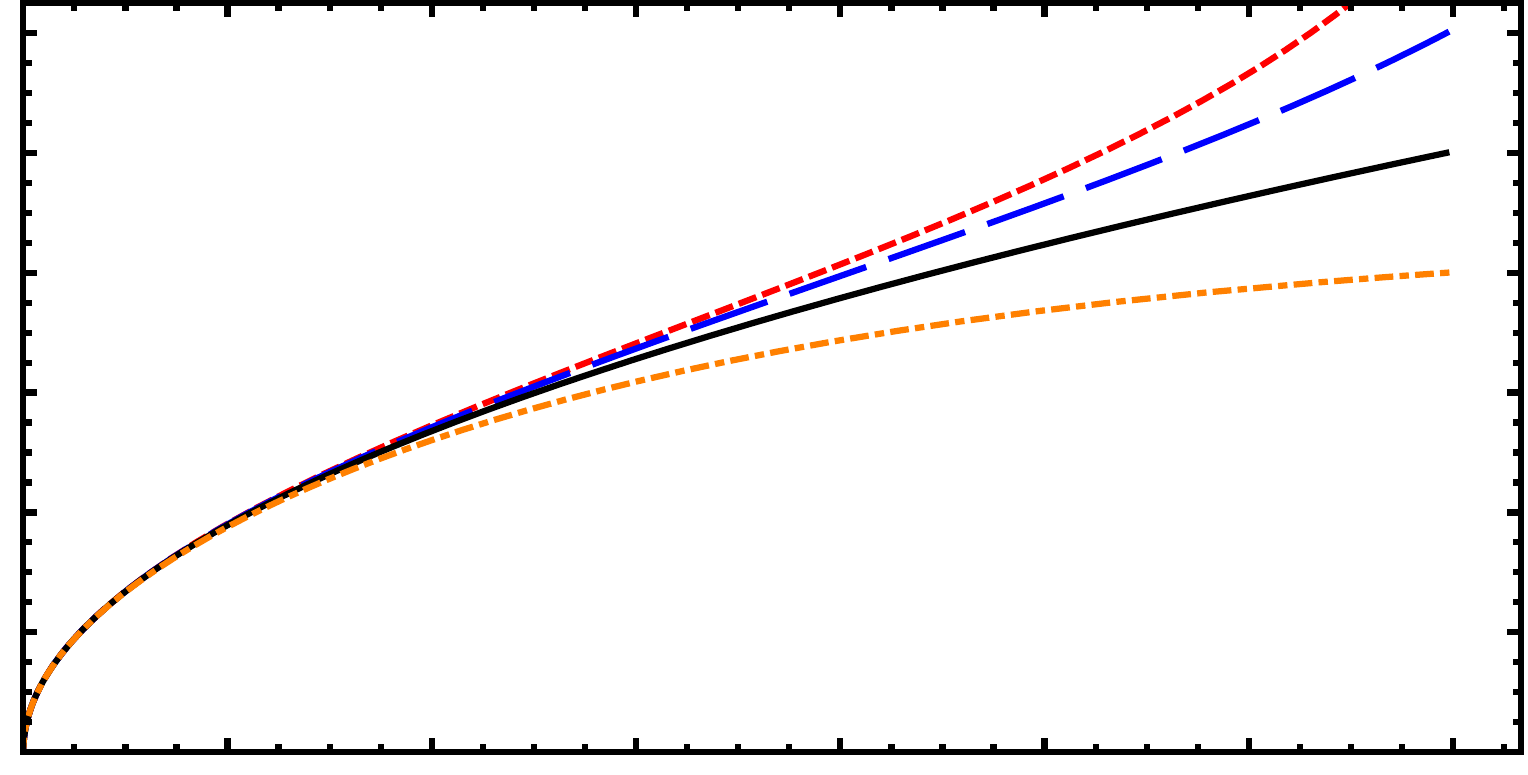}}%
  \end{picture}%
  
\caption{Numerical results for $R(N)$ for initial conditions close to the critical point $R_\text{in} \sim \sqrt{N_\text{in}}$ and $c = 0.01$.}
\label{fig:three_body_results}
\end{figure}

To complete the picture, we present numerical results for initial conditions close to criticality in Fig.\ \ref{fig:three_body_results}. As expected, we observe that all solutions flow towards the critical solution with $q(N) \sim \sqrt{N}$.

Their peculiar properties in terms of stability and criticality make condensates that behave according to \eqref{eq:coll_sols} interesting objects on their own. Ultimately, however, we are interested in a condensate that could mimic the behavior of black holes. Their very short collapse times rule out condensates that deay via a three-body process. We therefore now focus on condensates whose dominant decay mechanism is given by the two-body process Fig.\ \ref{fig:twobody}.

\subsection{Two body decay}
\label{sec:twobody}

In the case of $c_{(3)} = 0$, the simple solutions (\ref{eq:coll_sols}) are absent. 
Nevertheless, we find solutions that \emph{approximately} display critical behavior.

In order to see this, we write $R$ as
\begin{equation}
R = R_c(N)\,.
\end{equation}
Plugging this into Eqs.\ (\ref{eq:qddot},\ref{eq:ndotm}) with $c_{(3)} = 0$ and demanding mutual consistency of the equations yields a differential equation for $R_c(N)$. 
In absence of an exact solution to this differential equation, we write the relation between $N$ and $R$ as a power series:
\begin{equation}
R_c(N) = \ell_P\sqrt{a_0 N}\left(1 + \sum_k a_k N^{-k}\right)\,.
\label{eq:qntwo}
\end{equation}
Expanding in powers of $1/N$, we find up to second order 
\begin{equation}
R_c (N) = \left(1 - \frac{5 c_{(2)}^2}{a_0^4 N^2}\right)\ell_P\sqrt{a_0 N}
\end{equation}
with $a_0 \equiv \frac{1}{\sqrt{2}}\left(\frac{3}{4}\right)^3$. The criticality condition is hence fulfilled
up to corrections of order $1/N^2$.

To lowest order in $1/N$, we find the solution for the particle number 
\begin{equation}
N(t) \sim N_0 \left(1 - \frac{3}{2}\frac{c_{(2)} t}{a_0^{5/2}N_0^{3/2} \ell_P}\right)^\frac{2}{3} + {\cal O}(1/N)\,.
\label{eq:nofttwo}
\end{equation}
Not surprisingly, if we were to trust this solution up to the point of complete evaporation, we would obtain an evaporation time $t \sim \ell_P N_i^{3/2} \sim R^3/\ell_P^2$.

Again, we test the stability of the criticality condition by considering 
\begin{equation}
R(t) = R_c(N(t)) + \delta{R}(t)\,,
\end{equation}
linearizing in $\delta{R}$ and including contributions up to order\footnote{Note that if we expanded to higher orders, source terms would appear that are due to \eqref{eq:qntwo} deviating from the exact solution to the equations of motion.} $1/N^2$.
We obtain the two eigenvalues
\begin{equation}
\lambda_\pm = \pm \frac{1}{\ell_P}\sqrt{\frac{2}{a_0 N}} - \frac{3 c_{(2)}}{2 a_0^{5/2} \ell_P N^{3/2}}\,.
\end{equation}
For large $N$, $\lambda_+$ is positive. The critical solutions are thus unstable. The leading Lyapunov coefficient is found to be $\lambda_L \sim 1/\ell_P\sqrt{N}$.

This result opens up a curious side track. As was shown in \cite{Dvali:2013vxa}, the existence of an instability in the Gross-Pitaevskii equation leads to generation of one-particle entanglement on a time scale $t \sim \lambda_L^{-1} \log{N}$. Using the above result, we obtain $t \sim R \log{R}$, implying that quantum correlations will become important on timescales of the order of the scrambling time of black holes.
This could provide a first hint towards a fast scrambling behavior of these kind of condensates. 

In the case of black holes, the solution should follow the critical behavior at least up to the point in which $1/N$ effects can become dominant, a time roughly of the order of Page's time \cite{Page:1993df}. In the language of gravitons, this is due to the maximal packing property \cite{Dvali:2011aa} of BH, which reflects the fact that it is impossible to localize energy in a volume smaller than its own Schwarzschild radius. This kind of behavior cannot be concluded from our analysis. This may be due to several reasons. For one, it may be that maximal packing is intimately related with entropy. In the condensate picture, exponential degeneracy is only expected to be present at the critical point. This could imply that trajectories that follow critical behavior are quantum mechanically favored. If the large entropy is
a unique feature of GR it can never be seen in simpler toy models. This may be tested through an analysis of ``classicalizing'' \cite{Dvali:2010jz} theories, which are expected to show similar behavior. Most probably, this would require applying our analysis to a theory with more powers of derivatives.

On the other hand, it is conceivable that the apparent absence of maximal packing is not due to the toy model, but instead due to the approximation in which we neglect higher order correlators in the Bogoliubov hierarchy. In our toy model, the instability implies a deviation from critical behavior that becomes ${\cal O}(1)$ on the same timescale as the deviation from classicality. On the other hand, this in turn implies that corrections to Eqs.\ (\ref{eq:qddot},\ref{eq:ndotm}) can become relevant. A conclusive answer requires further analysis.

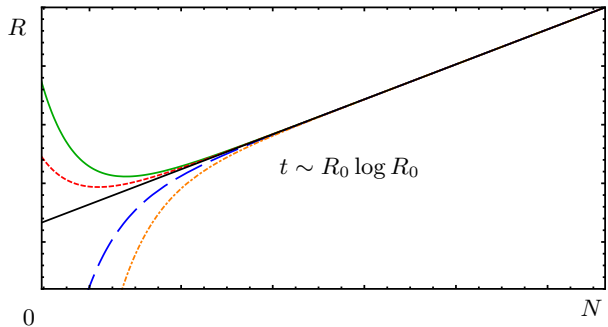
\begin{figure}[t]
  \setlength{\unitlength}{0.93\linewidth}
  \centering
  \begin{picture}(0.95,0.55)%
      \put(0.0,0.0){\makebox(0,0)[r]{\strut{} 0}}%
      \put(-0.03,0.48){\makebox(0,0){\strut{}$R$}}%
      \put(0.92,0.015){\makebox(0,0){\strut{}$N$}}%
      \put(0.52,0.25){\makebox(0,0){\strut{}$t \sim R_0 \log R_0$}}%
      \put(0.0,0.04){\includegraphics[width=0.88\linewidth]{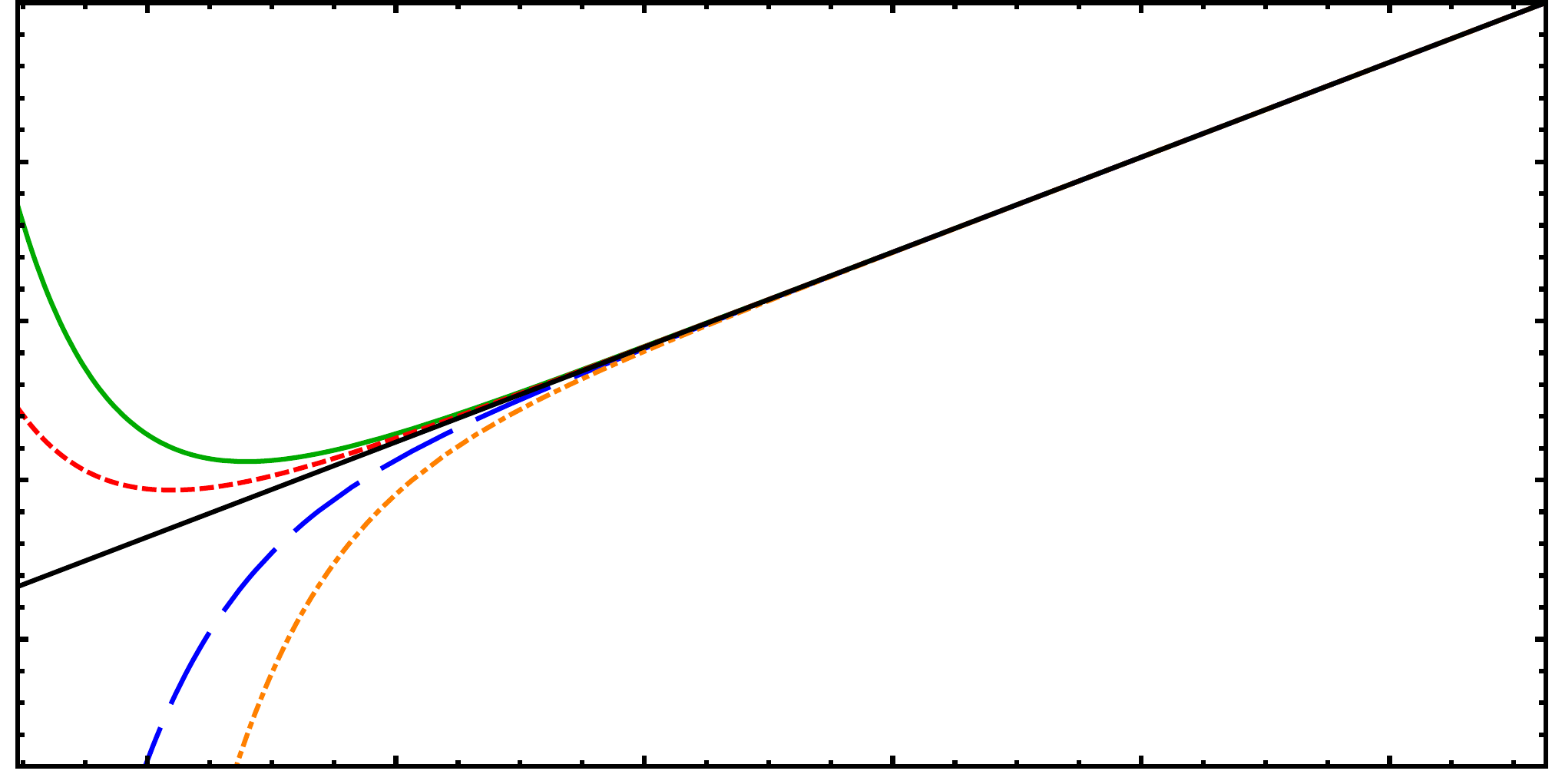}}%
  \end{picture}%
  
\caption{Numerical results for $R(N)$ for initial conditions close to the critical point $R_0 \sim \sqrt{N_0}$. The deviation from critical behavior becomes relevant at times $t \sim R_0 \log R_0$.}
\label{fig:two_body_results}
\end{figure}

For completeness, we show in Fig.\ \ref{fig:two_body_results} the behavior of the condensate for generic initial conditions in the vicinity of the critical solution. 
Indeed, we observe that after a time $t \sim R \log R$, the solutions diverge. For $R(t_\text{in}) > R_c(N(t_\text{in}))$ the condensate eventually ceases to collapse and turns around, while particle emission continues for a while. At some point, the condensate is so dilute that emission is effectively shut off and the particle number remains constant. On the other hand, for $R(t_\text{in}) < R_c(N(t_\text{in}))$, we observe a rapidly collapsing condensate after the instability time. The emission remains close to the solution \eqref{eq:nofttwo} until the condensate size has reduced to a fraction of its initial value. Then, almost all condensed particles are ejected within a very short time.

However, we stress here once again that due to the presence of the instability, we expect higher order correlations to become large around the instability time. The description in terms of an effective Gross-Pitaevskii equation is then likely to break down\footnote{Taking into account all corrections would lead to the well known BBGKY-hierarchy \cite{bbgky1,*bbgky2,*bbgky3,*bbgky4,*bbgky5}. In practice, the hierarchy has to be truncated at some order; in the case of rapidly growing correlations, this is difficult to do consistently. However, this is certainly a possible road that deserves further attention.}.

\section{Conclusions}

In this work, we have developed a toy model for Hawking evaporation in the context of the Bose condensate picture for black holes.
To this end, we have constructed a Hamiltonian that captures essential ingredients to the underlying physics of black hole evaporation, while at the same time being stripped down from some of the complications that arise in Einstein gravity. In particular, we have focused on a single degree of freedom and have turned off processes that violate particle number conservation.
We have introduced the Schwinger-Keldysh formalism for nonequilibrium dynamics of Bose condensates and derived the Keldysh action that describes a collapsing and evaporating condensate.

We have then chosen a variational approach to solve the ensuing equations of motion, using the number of condensed particles $N$ and the size of the condensate $R$ as variational parameters. The resultant action took the form of that of a relativistic particle in an effective $N$-dependent potential. In comparison with existing results in the literature, we have identified the corrections due to the relativistic dispersion relation.

Once the decay of the condensate was taken into account, we have found two possible behaviors. 
If the condensate is sufficiently inhomogeneous, 
evaporation is predominantly due to interactions in which one of the participants rescatters into the condensate.
In this case, the resulting decay rate is proportional to $\sqrt{N}/\ell_P$, leading to a lifetime of the bound state that is much shorter than what is expected for a black hole. 
Within the regime of validity of our assumptions, the three-body decay is required to be forbidden if the condensate should be a viable black hole candidate.

On the other hand, we have considered situations in which the three-body process is forbidden, for example by momentum conservation for a homogenous condensate, and decay can only occur through a two-body process in which both particles are ejected. In this case, we have discovered many of the properties that make a condensate picture for black holes appealing. Collapse and evaporation can happen nearly self-similarly, with width and particle number related via $R \sim \ell_P \sqrt{N}$ up to subleading $1/N$ corrections. The bound state then has a lifetime that may be as long as $\ell_P N^{3/2}$, much like semiclassical black holes. At the same time, the solution exhibits an instability of the form conjectured in \cite{Dvali:2013vxa} to be responsible for scrambling. The leading Lyapunov coefficient was found to be $1/\ell_P\sqrt{N}$, yielding a quantum break time of order $R \log R$, reminiscent of the scrambling time for black holes.

Many further advances are of course necessary before one can truly gauge whether the evolution of black holes may indeed be understood as the physics of Bose condensates. For one, further improvements of the toy model are in line to provide a better understanding of the processes in GR. This comprises the inclusion of particle number violating vertices and of the entire tower of interactions present in GR, as well as the generalization to non-vanishing helicity, and, in hand, the implementation of longitudinal modes that are responsible for the gravitational potential. Moreover, the compatibility of the presence of an instability with the continuously critical behavior that seems to be present for black holes needs to be understood. This is equivalent to a dynamical understanding of the ``maximal packing'' \cite{Dvali:2011aa} property of black holes, i.e.\ the fact that it is impossible to localize energy beyond its Schwarzschild radius. In the simplest case, steps in this direction may be taken already at the level of ``classicalizing'' \cite{Dvali:2010jz} scalar field theories. It may however well be that it is the specific structure of General Relativity that holds the key to this behavior.

\section*{Acknowledgments}
We wish to thank Sarah Folkerts for her important contributions during early stages of this work. Moreover we thank Rembert Duine, Gia Dvali, Daniel Flassig, Stefan Hofmann, Florian K\"uhnel, Alexander Pritzel, Tehseen Rug, Henk Stoof and Bo Sundborg 
for important comments and discussions.
The work of N.W.\ was supported in part by the Humboldt Foundation and by the Swedish Research Council (VR) through the Oskar Klein Centre. 

\appendix*

\section{}
\begin{figure}[b]
	\setlength{\unitlength}{0.93\linewidth}
	\centering
	\begin{picture}(0.95,0.61)%
	\put(0.05,0.05){\makebox(0,0)[r]{\strut{}$N-n$}}%
	\put(0.0,0.26){\makebox(0,0){\strut{}$n$}}%
	\put(0.95,0.15){\makebox(0,0){\strut{}$N-k$}}%
	\put(0.85,0.5){\makebox(0,0){\strut{}$k'$}}%
	\put(0.06,0.0){\includegraphics[width=0.75\linewidth]{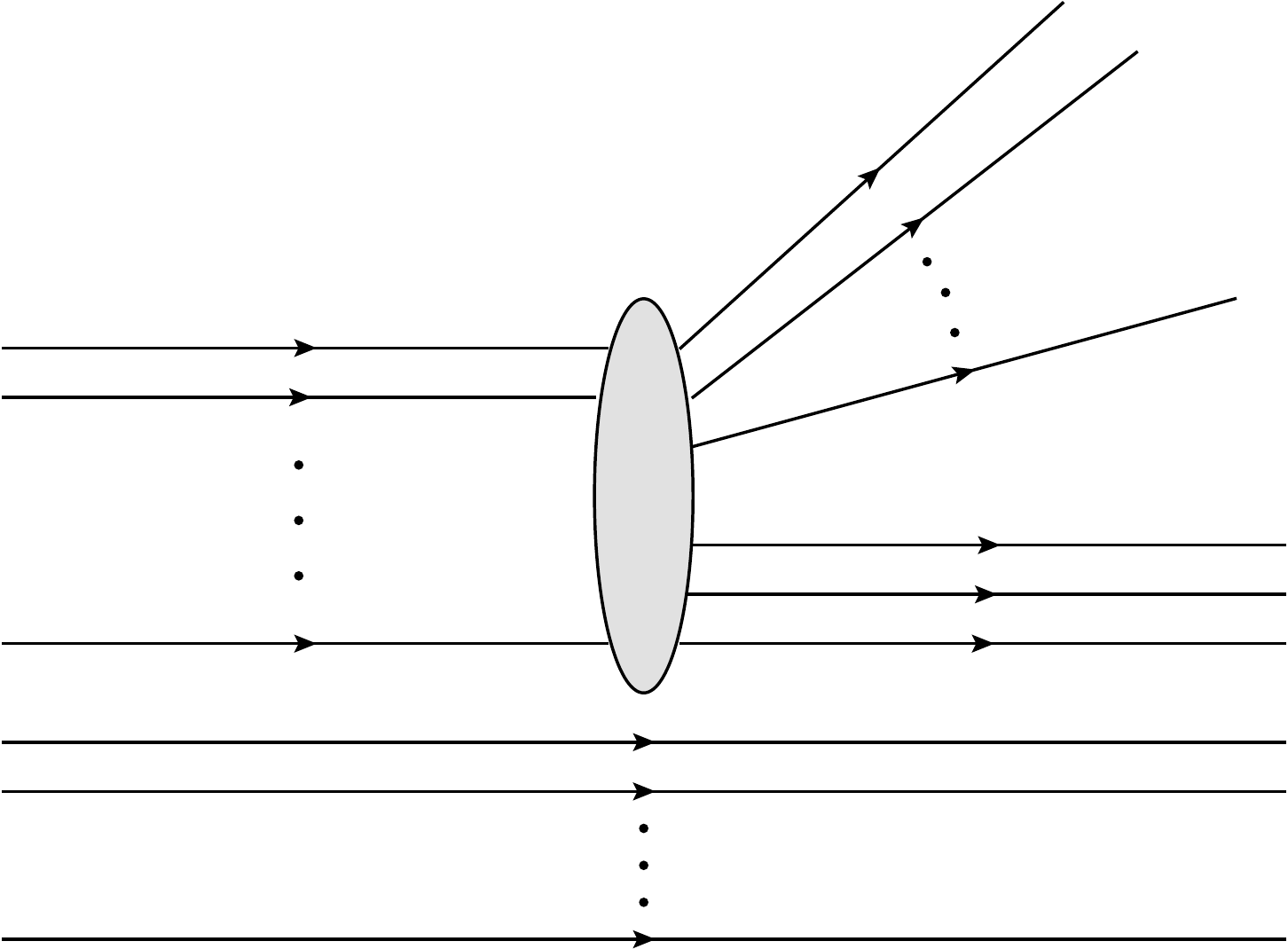}}%
	\end{picture}
	\caption{Generic diagram for transition $N \to N-k$. $n$ constituents interact, leading to the emission of $k'$ particles.}
	\label{fig:many_decay}
\end{figure}
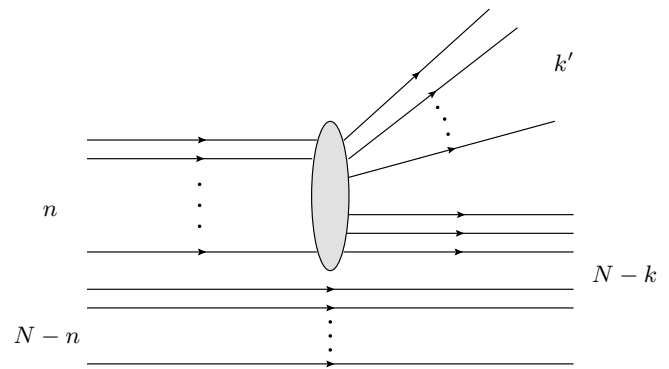

In this appendix, we briefly revisit the initial assumptions stated in section \ref{proto}. Using our results from section \ref{sec:sols}, we can study their viability by estimating a generic decay $N \to N - k$. The setup is illustrated in Fig.\ \ref{fig:many_decay}. 
Due to interaction of $n$ constituents, $k'$ particles are emitted, while $n - k$ particles reenter the condensate. The grey ellipse represents a generic tree-level interaction, yielding $2n-k+k'-2$ powers of the Planck mass.
We require that no on-shell fluctuations rescatter on the condensate, since such processes receive additional suppression factors of $1/N$.
Under the assumption that the individual momentum transfer scales as $N^{-1/2}$ (which is the case as long as $n, k, k' \ll N$), we obtain for the squared matrix element
\begin{equation}
|{\cal M}|^2 \sim \frac{N!}{(N-n)!} N^{-2n+k-k'+2}\frac{(N-k)!}{(N-n)!}
\end{equation}
which we may approximate by
\begin{multline}
|{\cal M}|^2 \sim \left(1-\frac{2n-k}{N}\right)^N e^{-2n + k} \\ \left(1 - \frac{2n^2 - n + k^2 - k/2}{N} \right) N^{2-k'}.
\end{multline}
The squared amplitude is thus bounded by
\begin{equation}
\label{eq:bound}
|{\cal M}|^2 < N^{2-k'}.
\end{equation}
We see that the largest contribution indeed stems from the lowest order vertex. Moreover, the scaling from section \ref{sec:sols} survives.

Note that \eqref{eq:bound} is a crude upper bound on the scaling of the higher order interactions. More precise answers, in particular on the resummed rate, require an in-depth analysis of the specific vertex structure of the theory; in the case of GR, additional suppression factors are found \cite{Dvali:2014ila}.

\bibliography{biblio}

\end{document}